# *The EXoplanetary Circumstellar Environments and Disk Explorer (EXCEDE)*

Olivier Guyon*[a], Glenn Schneider[a], Ruslan Belikov[b], Domenick J. Tenerelli[c]
[a]Steward Observatory, University of Arizona, 933 Cherry Ave., Tucson, AZ 85721;
[b]NASA Ames Research Center, Moffet Field, Mountain View, CA 94035, USA;
[c]Lockheed Martin Space Systems Company, Palo Alto, CA

## ABSTRACT

We present an overview of the EXoplanetary Circumstellar Environments and Disk Explorer (*EXCEDE*), selected by NASA for technology development and maturation. *EXCEDE* will study the formation, evolution and architectures of exoplanetary systems, and characterize circumstellar environments into stellar habitable zones. *EXCEDE* provides contrast-limited scattered-light detection sensitivities ~ 1000x greater than HST or JWST coronagraphs at a much smaller effective inner working angle (IWA), thus enabling the exploration and characterization of exoplanetary circumstellar disks in currently inaccessible domains. *EXCEDE* will utilize a laboratory demonstrated high-performance Phase Induced Amplitude Apodized Coronagraph (PIAA-C) integrated with a 70 cm diameter unobscured aperture visible light telescope. The *EXCEDE* PIAA-C will deliver star-to-disk augmented image contrasts of < 10E-8 and a 1.2 $\lambda$/D IWA or 0.14" with a wavefront control system utilizing a 2000-element MEMS DM and fast steering mirror. *EXCEDE* will provide 0.12" spatial resolution at 0.4 μm with dust detection sensitivity to levels of a few tens of zodis with two-band imaging polarimetry. *EXCEDE* is a science-driven technology pathfinder that will advance our understanding of the formation and evolution of exoplanetary systems, placing our solar system in broader astrophysical context, and will demonstrate the high contrast technologies required for larger-scale follow-on and multi-wavelength investigations on the road to finding and characterizing exo-Earths in the years ahead.

## 1. INTRODUCTION

The EXoplanetary Circumstellar Environments and Disk Explorer1 (*EXCEDE*), recently selected by NASA for technology development, uses a 0.7 m diameter off-axis telescope to directly image the immediate surrounding of nearby stars at high contrast. *EXCEDE* delivers high contrast images thanks to a starlight suppression system (SSS) including a coronagraph and active wavefront control. *EXCEDE* will operate in visible light, and provide dual-band polarimetric images of circumstellar disks. While *EXCEDE* is primarily aimed at imaging disks, it will also be able to image a few giant exoplanets in reflected light. In its 3-yr nominal mission duration, *EXCEDE* will observe 350 targets from a Sun-synchronous low earth orbit.

*EXCEDE*'s telescope diameter is contrainted by mission cost at 70 cm, but its use of a high efficiency low inner working angle (IWA) PIAA-type coronagraph[2-10], and its ability to image disks at short optical wavelength, mitigates its small telescope aperture. It can thus detect and characterize disks that are both fainter and closer in that possible with current ground-based or space-based facilities, providing a rich science return, as described in section 2. The instrument design is detailed in section 3, and incorporates coronagraphic and wavefront control techniques which have been developed to support future NASA high contrast imaging missions. In addition to fullfilling a set of science objectives detailed in section 2, *EXCEDE* is therefore also a technology precursor to future more capable high contrast imaging missions that will be able to detect and characterize habitable planets around nearby stars. The mission implementation is described in section 4.

*guyon@naoj.org; phone 1 818 293 8826

## 2. SCIENCE GOALS & OBJECTIVES

### 2.1 Science goals

Thanks to its small IWA, *EXCEDE* will direcly image the habitable zones of a statistically significant number of stars, and will therefore **characterize the circumstellar environments in habitable zones and assess the potential for habitable planets**. While not being able to directly observe such planets, *EXCEDE* will contrain the planetary architecture in the habitable zone thanks to planet-dust interactions.

*EXCEDE*'s observations of circumstellar protoplanetary and debris disks around a large sample of stars will enable **understanding the formation, evolution, and architecture of planetary systems**.

In addition to *EXCEDE*'s science goals listed above, *EXCEDE* will **develop and demonstrate advanced coronagraphy in space, enabling future exoplanet imaging missions**. *EXCEDE* is therefore of high importance for the longer term NASA goal of finding life outside our solar system.

### 2.2 Science Objectives

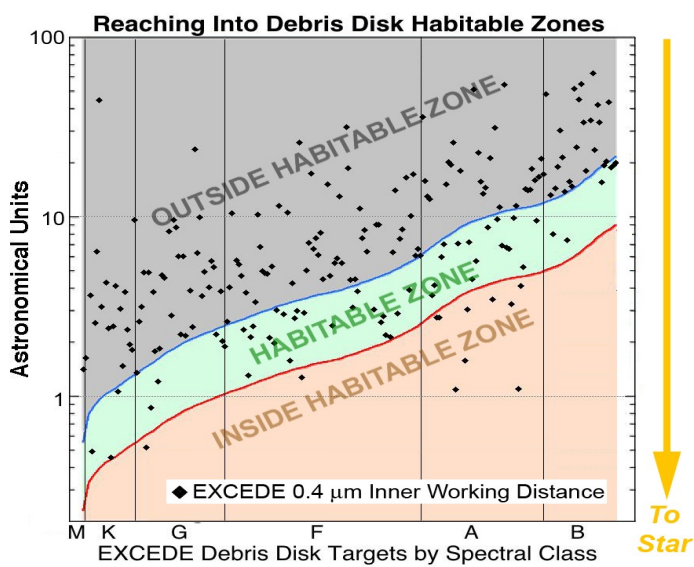

Figure 1. EXCEDE's inner working angle at 0.4 µm is shown here in astronomical units (vertical axis) for each target. Targets are ordered according to stellar type (horizontal axis) and the habitable zone is overlayed. For targets falling in the central green band of this figure, EXCEDE will image part of the habitable zone. For targets falling in the red lower part of the figure, EXCEDE will map dust in the entire habitable zone, while it will lack the angular resolution to access the habitable zone for targets in the upper part of this figure.

*EXCEDE* will **explore the amount of dust in Habitable Zones** (where dust indirectly traces the level of terrestrial planet bombardment by asteroids and meteorids). *EXCEDE* will provide direct images of scattered light debris disks around a sample of ~ 230 nearby (≤ 100 pc) stars revealing the levels of zodiacal light (ZL) present in these systems. ZL is a proxy for the richness of planetesimal belts and their degree of gravitational stirring, and is an indirect indication of the level of bombardment that might be experienced by terrestrial planets in these systems. As shown in figure 1, for more than 1/4 of the debris disk targets, *EXCEDE*'s 0.14" IWA enables spatially resolved imaging in the circumstellar HZ where liquid water can exist on planetary surfaces. *EXCEDE* imagery will probe far interior to the Kuiper belt regions of nearby stars, into the now-elusive terrestrial planet and habitable zones of these dusty planetary systems, providing evidence for asteroid belts, comets and unseen planets.

*EXCEDE* will help determine **if this dust will interfere with future planet-finding missions.** The amount of dust in HZs is key to determining the best strategies to image Earth-like exoplanets, as dust-scattered starlight is the main source of astrophysical "noise" in detecting such faint point sources. *EXCEDE* will have the sensitivity to detect ZL at the 100 zodi level for nearby stars.

*EXCEDE* will constrain the **composition of material delivered to planets**. Identifying the presence of icy and organic-rich disk grains will give the first clues to the presence of volatiles important for life. *EXCEDE*'s two-band imaging polarimeter is crucial to disentangling the dynamical and compositional history of disks.

*EXCEDE* will investigate what **fraction of systems have massive planets on large orbits**. *EXCEDE*'s image contrast and 120 mas spatial resolution (e.g., 1.2 AU at 10 pc) will vastly increase the number of Neptune-analogs discovered from dynamical influences on debris disks. Structure in disks betrays the presence of planetary systems. *EXCEDE* will reveal the radial locations of planetesimal belts – a powerful indicator of gas-giant and ice-giant planets.

*EXCEDE* will observe **how protoplanetary disks make Solar System-like architectures**. *EXCEDE* images will reveal disk sub-structures including large (> 20 AU) cavities and gaps associated with young Jovian-mass images. *EXCEDE* will observationally test models that predict gaps opening in circumstellar disks as a result of tidal interactions with giant planets. Theoretical models predict material-depleted disk "gaps", observable with *EXCEDE*, evolving over time due to the presence of co-orbiting planets

*EXCEDE* will measure the **reflectivity of giant planets and constrain their compositions**. *EXCEDE* will produce the first images of extrasolar planets in the inner (0.5 < a < 7 AU) regions of mature planetary systems like our own. Simultaneous measurements of degree of polarization, color, and total brightness will probe the atmospheric compositions of cool extra-solar giant planets for the first time.

## 3. INSTRUMENT DESIGN

### 3.1 Overview

*EXCEDE* delivers high contrast images thanks to its starlight suppression system (SSS), composed of a coronagraph and fine wavefront control hardware and software. *EXCEDE*'s instrument is designed to provide high contrast imaging of the immediate surrounding of stars in two spectral bands centered around 0.4 μm and 0.8 μm respectively. Switching between the two bands is done with a filter wheel, so there is a single SSS channel. A block diagram of the SSS is shown in figure 2. The coronagraph design is described in section 3.2, and the wavefront control hardware and software are described in section 3.3. Expected performance for the *EXCEDE* SSS is described in section 3.4.

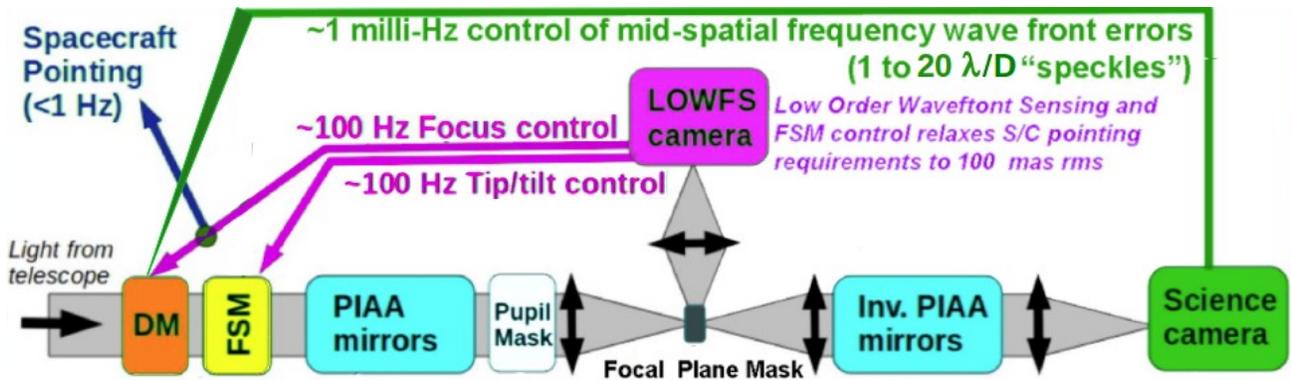

Figure 2. EXCEDE starlight suppression system overview. The coronagraph, described in section 3.2, uses lossless apodization performed by PIAA optics. Wavefront control, described in section 3.3, uses the focal plane image to sense scattered light and a dedicated coronagraphic low-order wavefront sensor for measuring pointing errors. A 2000-actuator deformable mirror (DM) and a fine steering mirror (FSM) perform the required wavefront corrections.

### 3.2 Coronagraph

EXCEDE's coronagraph uses the Phase-Induced Amplitude Apodization (PIAA) technique, a high efficiency alternative to conventional apodization. Conventional apodization relies of selective transmission with a pupil plane mask – a robust but low efficiency solution. The coronagraph design must then trade contrast against efficiency and inner working angle (IWA), a particularly painful choice for high contrast imaging instruments/missions.

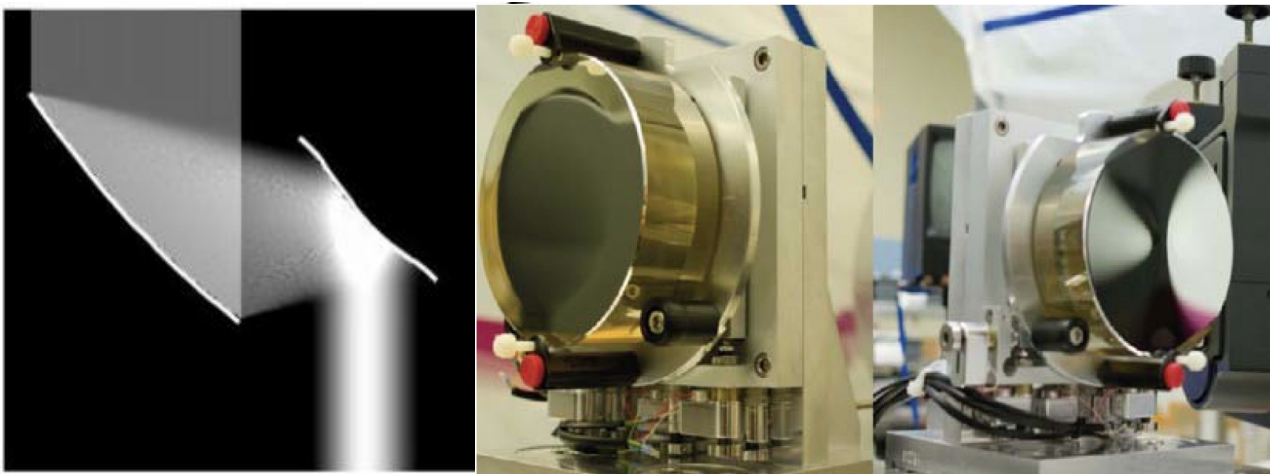

Figure 3. PIAA principle (left): the telescope beam is reshaped into an apodized beam using aspheric optics, with no loss in throughput or angular resolution. Aspheric PIAA mirrors, manufactured by L3-Tinsley, are shown (center: PIAA M1; right: PIAA M2).

Phase-Induced Amplitude Apodization (PIAA) achieves the same apodization with beam shaping, using aspheric optics such as the ones shown in figure 3 to reshape the telescope beam into an apodized beam with no loss in throughput or angular resolution[ref]. This coronagraphic approach offers very high performance, as it combines full throughput, small IWA and uncompromized angular resolution. EXCEDE uses reflective PIAA optics (aspheric mirrors) similar to the mirrors shown on figure 3, thus producing an achromaticity apodization. The challenging manufacturing of the aspheric optics is mitigated by adopting a ``hybrid'' approach where apodization is shared between the PIAA mirrors and a mild apodizer (which does not incurr a significant throughput loss), located after the PIAA mirrors.

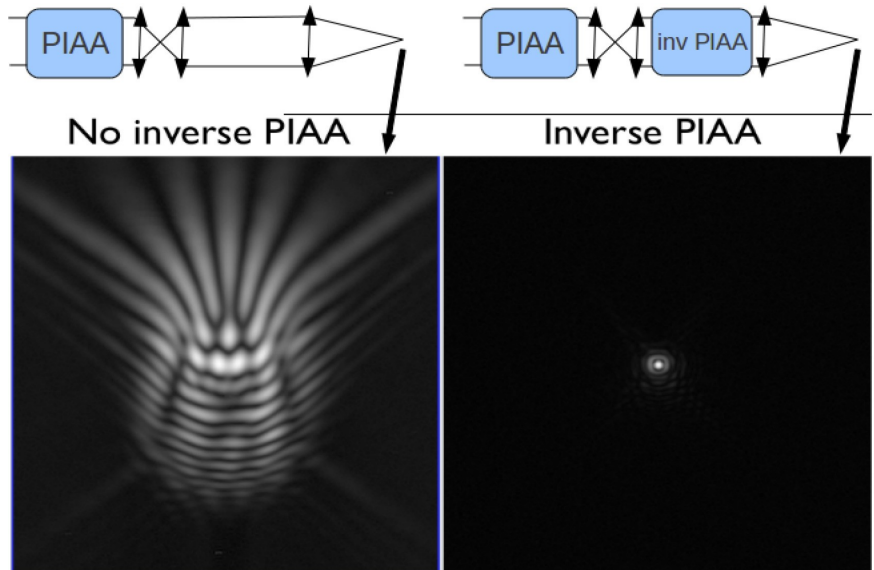

Figure 4. Effect of the inverse PIAA on off-axis image quality, as demonstrated by laboratory images obtained for the Subaru Coronagraphic Extreme-AO (SCExAO) system. Left: off-axis PSF without the inverse PIAA. Right: off-axis PSF wih the inverse PIAA. For both images, the sources is located 20 λ/D from the optical axis defined by the PIAA optics.

EXCEDE's focal plane mask is circular and opaque (in fact partially reflective, as will be described in section 3.3), and the size of the focal plane mask is slightly undersized following the same design approach as the Apodized Pupil Lyot Coronagraph (APLC)[ref]. Reducing the focal plane mask offers two key advantages: the apodization becomes weaker, resulting in easier to manufacture PIAA optics, and the coronagraph IWA is reduced. In this design, the focal plane mask size is however critical, and can only be matched to the apodization profile at a single wavelength. EXCEDE's PIAA was designed to provide as small an IWA as possible at 1e-6 raw contrast in a 20% wide band, ledading to a 1.2 $\lambda/D$ IWA (IWA is defined here as the separation for which the throughput reaches 50% of the nominal coronagraph throughput measured at large angular separations).

A set of inverse PIAA optics is also required at the back end of the coronagraph to cancel field aberrations introduced by the first set of PIAA optics[11]. This inverse set plays no role in the coronagraphic process, but considerably extends the field of view over which the PSF is diffraction limited. Laboratory images obtained with and without the inverse PIAA for a source 20 $\lambda/D$ IWA from the optical axis are shown in figure 4, demonstrating the inverse PIAA's effect on image quality. EXCEDE's inverse PIAA optics are mirrors of the same size as the direct PIAA optics.

### 3.3 Wavefront Control: Pointing

Fine pointing control is essential to maintain EXCEDE's contrast performance at its small IWA. The pointing requirement for EXCEDE is 2 mas RMS at the coronagraph input (light entering the PIAA optics) in order to maintain a 1e-6 contrast at 1.2 $\lambda/D$. EXCEDE achieves this pointing requirement by using the Coronagraphic Low-Order Wavefront Sensor[12] (CLOWFS) technique to accurately measure pointing errors within the coronagraph, as shown in figure 5. The focal plane mask is partially reflective: light falling between 0.6 $\lambda/D$ and 1.2 $\lambda/D$ from the optical axis is reflected to a pointing camera. The camera sees a defocused image of the focal plane mask. Small changes in instrument pointing produce a macroscopic change in the camera image, which is a linear function of the pointing error. The pointing offset signal is therefore obtained by linear decomposition of the CLOWFS camera image.

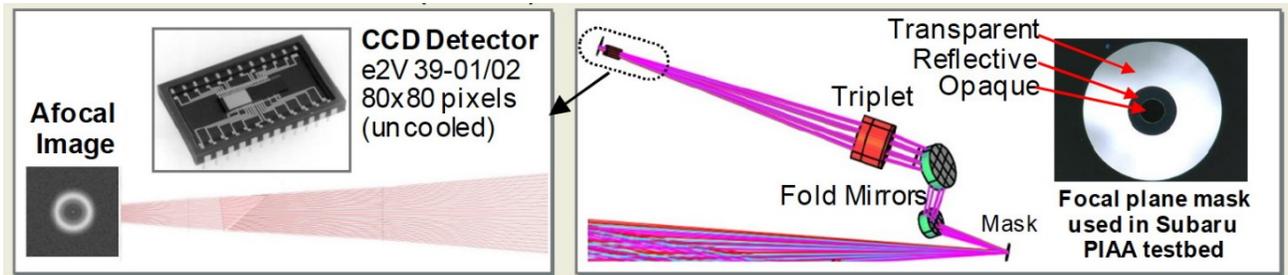

Figure 5. EXCEDE'scoronagraphic low-order wavefront sensor (CLOWF) uses light reflected by the coronagraphic focal plane mask (right) to measure pointing errors. The reflected light, which originates from an annulus extending from 0.6 $\lambda/D$ to 1.2 $\lambda/D$ radius, is re-imaged on a 80x80 pixel CCD (left). The image is slightly defocused to allow unambiguous measurement of focus as well as pointing.

EXCEDE's CLOWFS measures the instrument pointing to the required 0.01 $\lambda/D$ precision in 3.5 ms for targets brighter than $m_V$ = 10. A fast frame readout CCD camera is used as the CLOWFS sensor, and a fast steering mirror (FSM) located before the PIAA optics is actively driven to maintain instrument pointing. This control loop runs at ~100 Hz . The FSM tip-tilt is slowly offloaded to the telescope pointing to avoid FSM actuator saturation and to avoid large beam walk within the optical train. EXCEDE's CLOWFS also measures instrument focus, which is offloaded to the SSS deformable mirror described in the next section.

### 3.4 Wavefront Control: Focal Plane Speckle Control

EXCEDE measures and controls scattered light in the focal plane with a single 2000-actuator MEMS type deformable mirror located ahead of the PIAA optics. The complex amplitude of focal plane speckles is derived from intensity modulation obtained by adding known complex amplitude speckles and observing the coherent interference between the unknown speckle to be measured and the known probles introduced. This scheme[13] is now routinely used in high contrast laboratories, and has been demonstrated to contrast levels well beyond EXCEDE's requirement .

With a single deformable mirror, EXCEDE is not able to fully compensate phase and amplitude wavefront errors over a 20% spectral band. Modeling however shows that a single deformable mirror is sufficient to achieve 1e-6 raw contrast from 1.2 λ/D to 2 λ/D, and 1e-7 raw contrast from 2 λ/D to 20 λ/D, as shown in the next section.

**3.5 Expected Contrast**

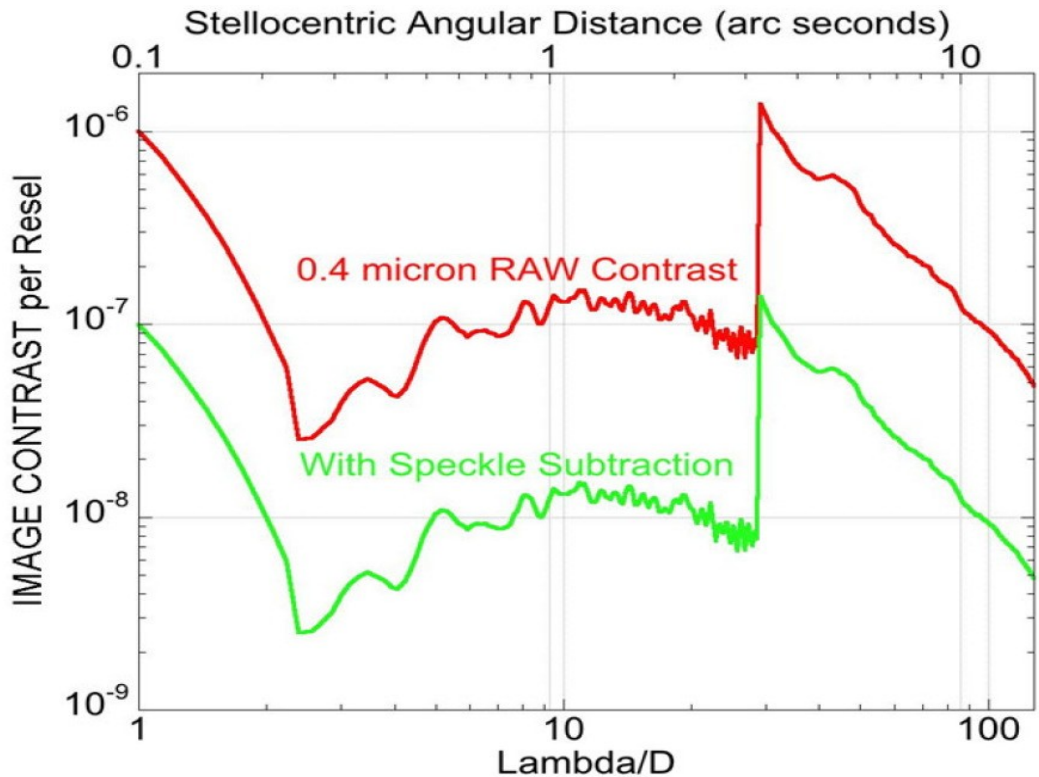

Figure 6. Simulated contrast performance for EXCEDE at 0.4 μm. See text for details.

A model of the optical train, including surface errors on each optical element, has been developed to estimate EXCEDE's contrast performance. A power spectral density (PSD) of surface errors representative of small size powered optics, provided from L3-Tinsley, was used for all mirrors except the highly aspheric PIAA mirrors. Actual wavefront measurements performed on the PIAA optics shown in figure 3 were used in the model. Several wavefront control modes and assumptions were explored. It was assumed, somewhat arbitrarily, that calibration techniques (differential imaging, PSF subtraction) will calibrate scattered light to $1/10^{th}$ of its raw level. This assumption will need to be revisited and refined with a dynamical model of the instrument/instrument to accurately take into account temporal variation of the wavefont.

The expected raw contrast is shown in figure 6 in the default EXCEDE observation mode, for which the MEMS DM is correcting phase-only aberrations, and therefore attempting to create a full size dark hole extending 360 deg around the optical axis. The raw contrast at small angular separations (up to ~3 λ/D radius) is limited by chromaticity in the coronagraph design, while the raw contrast further out is limited by uncorrected amplitude effects that originate from surface errors on optics non conjugated to the DM. The expected raw contrast is about 1e-6 at the coronagraph's IWA (1.2 λ/D) and is about 1e-7 outward of 2 λ/D and up to the outer working angle defined by the DM actuator pitch.

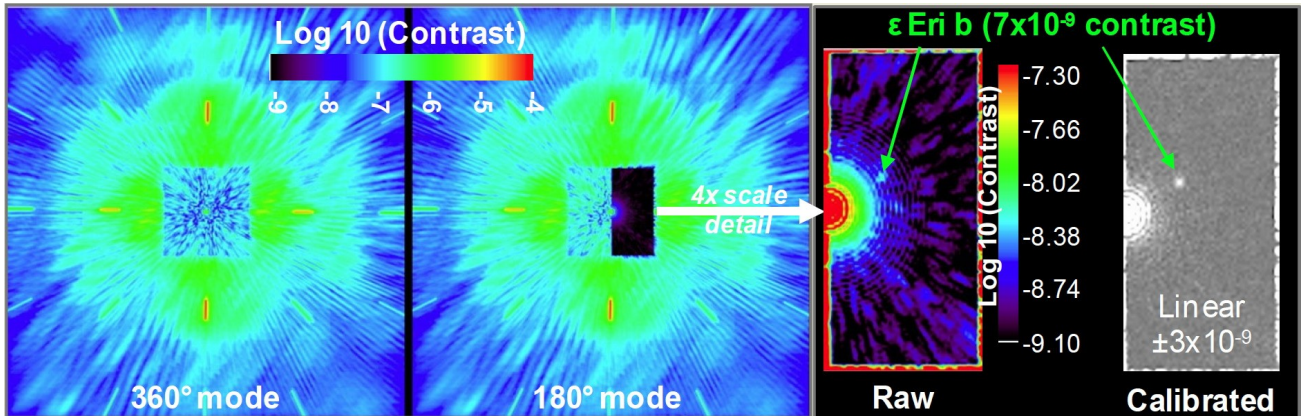

Figure 7. Simulated EXCEDE PSFs in 360 deg and 180 deg high contrast modes (left, center). The 180 deg mode offers deeper contrast, allowing the direct imaging of extra-solar giant planets, as illustrated by the simulated image of the ε Eri b planet (right).

EXCEDE's wavefront control system can also operate in 180 deg mode, attempting to suppress scattered light on only one side of the focal plane. In this mode, both amplitude and phase errors can be compensated, and the contrast achieved is therefore deeper, only limited by chromaticity effects due to diffractive propagation through the optical train and wavefront temporal stability. Figure 7 shows simulated EXCEDE PSFs in both modes. The 180 deg mode will be used for direct imaging of extrasolar giant planets, while the 360 deg mode is the default observing mode for disk observations.

Thanks to its small IWA coronagraph, high contrast, and short visible wavelength, EXCEDE will image disks that are significantly fainter than those at the detection limit of current instruments. Figure 8 shows simulated EXCEDE images of two disks that have been imaged by Hubble Space Telescope (HST) at near-IR wavelength, illustrating EXCEDE's sensitivity and ability to image faint disks at small angular separation. EXCEDE's angular resolution at 0.4 µm (118 mas) is similar to HST's angular resolution at 1.1 µm (95 mas), but its sensitivity is better, and its coronagraph can reach much closer in.

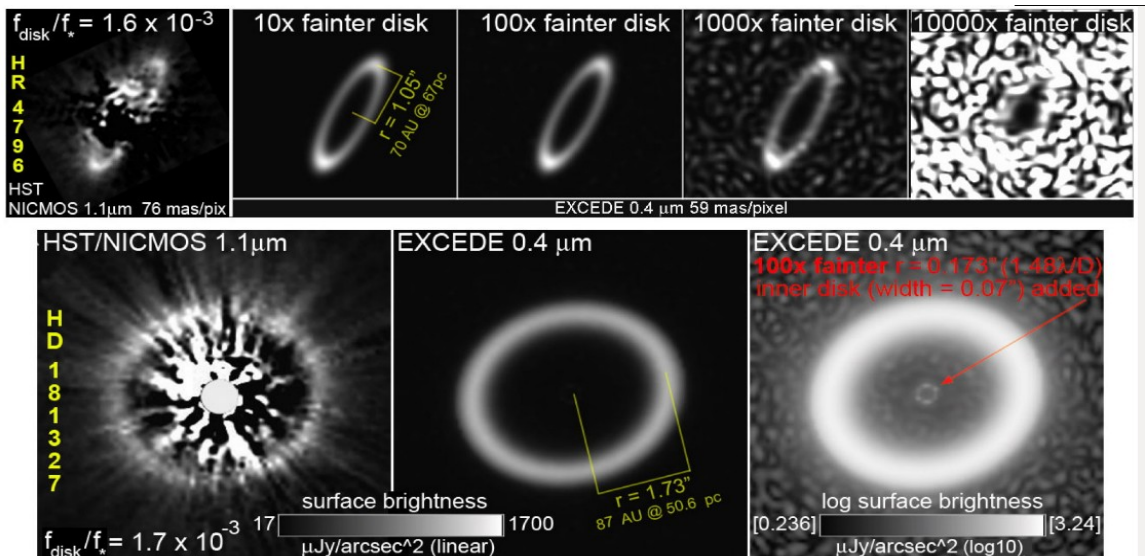

Figure 8. Simulated EXCEDE observations of known debris disks, compared to HST images.

## 4. MISSION IMPLEMENTATION

*EXCEDE* will be on a sun-synchronous low-earth orbit, at 2000 km altitude. The nominal mission duration is 3 yr, during which 350 targets will be observed.

### 4.1 Telescope

*EXCEDE* uses a 0.7 m diameter off-axis telescope, shown in figure 9. The starlight suppression module is mounted on the side of the telescope, near the secondary mirror. The science instrument module (telescope + SSS) is relatively compact, with a total length of under 2.7 m. The telescope primary and secondary mirrors produce a collimated beam which is then compressed to 20mm diameter by two additional mirrors (OAP-1 and beam compressor secondary).

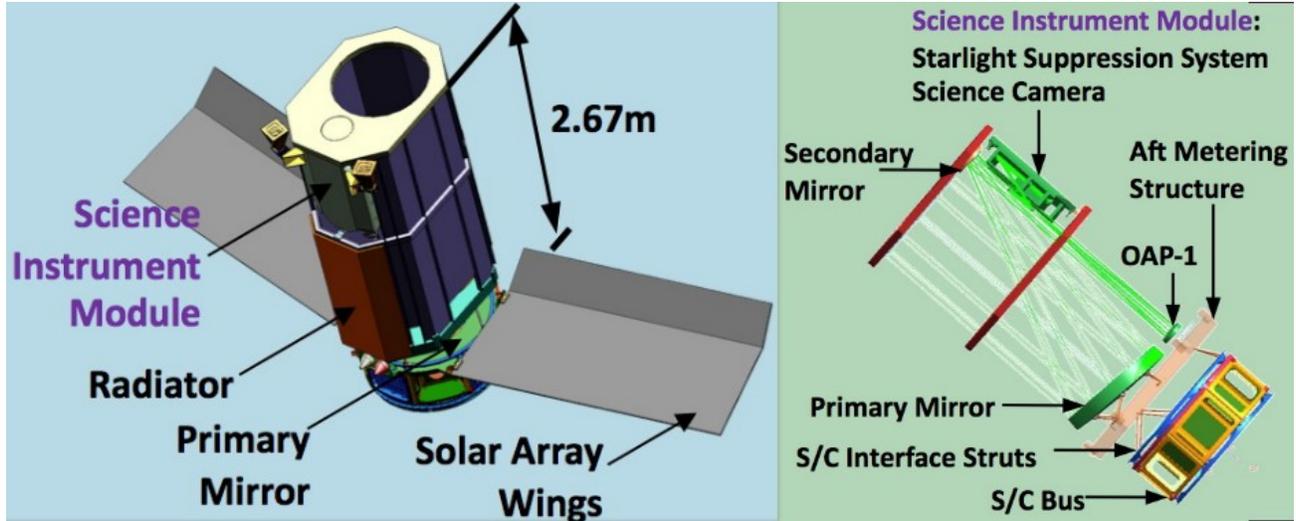

Figure 9. EXCEDE telescope and spacecraft configuration. The starlight suppression module is located on the side of the telescope tube.

### 4.2 Starlight Suppression System

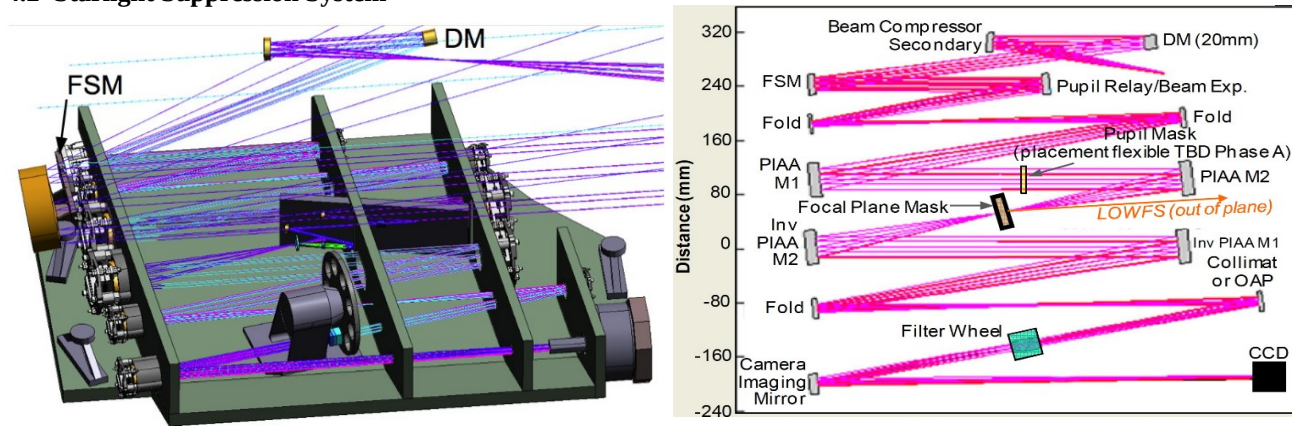

Figure 10. EXCEDE's starlight suppression system (SSS). The 3D drawing (left) shows the location of elements on the optical bench. Right: top view of the SSS.

EXCEDE's starlight suppression system (SSS), shown in figure 10, is mounted on the side of the telescope tube. Its elements are located on a single optical bench, preferentially along two rows between which the light bounces back and forth. The CCD focal plane array is located at one of the corners of the SSS. One of the largest elements, clearly visible on figure 10, is the filter wheel, which is located downstream of the coronagraph optics. As shown in figure 11, it contains the filters and Wollaston prisms necessary for target acquisition, dark acquisition and science observations with and without dual polarimetric imaging in both the 0.4 μm and 0.8 μm bands.

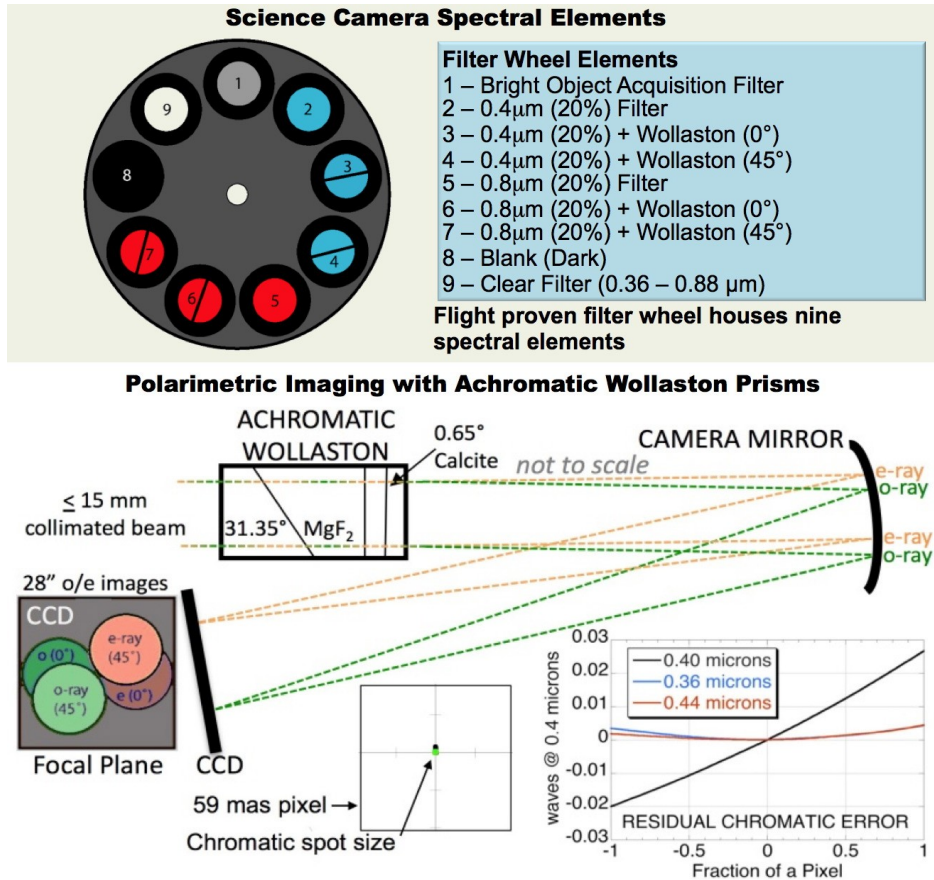

Figure 11. EXCEDE's camera design, showing elements at the end of the optical train: filter wheel containing filters and Wollaston prisms, camera mirror, and focal plane array.

## 5. CONCLUSIONS

EXCEDE, despite its moderate telescope size (0.7 m diameter) offers unprecedented sensitivity to circumstellar disks, thanks to its high efficiency low IWA PIAA-based coronagraph, its wavefront control system and its short observation wavelength. EXCEDE will image and characterize a large number of circumstellar disks, and, for the most favorable targets, its angular resolution and IWA will allow it to observe them in the habitable zone. EXCEDE will also be able to directly image a few extrasolar giant planets.

EXCEDE is an important step towards future more ambitious direct imaging mission(s) aimed at direct imaging and spectroscopic characterization of habitable planets. First, EXCEDE will measure the amount of exozodiacal light – the dominant source of astrophysical noise when imaging small rocky planets – around potential targets for such a future mission. Second, EXCEDE will validate key wavefront control and coronagraphy technologies/hardware that are essential for future high contrast imaging missions.

The EXCEDE team is now focusing its efforts on validating key technologies in the laboratory, demonstrating with an EXCEDE-like instrument configuration that the raw contrast level required for EXCEDE can be met. This effort is described in a separate paper[14]. The EXCEDE SSS architecture is also similar to the Subaru Coronagraphic Extreme-AO instrument deployed on the ground-based 8.2 m Subaru Telescope, and described in a separate paper[15].